\documentclass[universe,article,accept,pdftex,moreauthors]{Definitions/mdpi} 
\firstpage{1} 
\pubvolume{1}
\issuenum{1}
\articlenumber{0}
\pubyear{2024}
\copyrightyear{2024}
\datereceived{2023 December 14}
\daterevised{2024 January 7} 
\dateaccepted{2024 January 12} 
\datepublished{2024 January } 
\hreflink{https://doi.org/} 

\usepackage{natbib}

\newcommand{\oiii}{\textrm{[O\,{\footnotesize III}]}}
\newcommand{\feii}{\textrm{Fe}\,{\footnotesize II}}
\newcommand{\kms}{km\,s$^{-1}$}

\Title{Changing-look NLS1 galaxies, their detection with SVOM, and the case of NGC 1566}

\TitleCitation{Changing-look NLS1 galaxies, their detection with SVOM, and the case of NGC 1566}

 \Author{D. W. Xu $^{1,2}$*, S. Komossa$^{3,1}$*, D. Grupe$^{4}$, J. Wang$^{5,1}$, L. P. Xin$^{1}$, X. H. Han$^{1}$, J. Y. Wei$^{1,2}$, J. Y. Bai$^{1}$, E. Bon$^{6}$, F. Cangemi$^{7,8}$, B. Cordier$^{9}$, M. Dennefeld$^{10}$, L. C. Gallo$^{11}$, W. Kollatschny$^{12}$, De-Feng Kong$^{5}$, M. W. Ochmann$^{12,13}$, Y. L. Qiu$^{1}$, N. Schartel$^{14}$ }

\AuthorNames{D. W. Xu, S. Komossa, D. Grupe, et al.}

\AuthorCitation{D. W. Xu, S. Komossa, D. Grupe, et al.}

\address{%
$^{1}$ \quad Key Laboratory of Space Astronomy and Technology, National Astronomical Observatories, Chinese Academy of Science, Beijing 100101, China\\
$^{2}$ \quad School of Astronomy and Space Science, University of Chinese Academy of Sciences, Beijing 100049, China\\
$^{3}$ \quad Max-Planck Institut f\"ur Radioastronomie, Auf dem H\"ugel 69, 53121 Bonn, Germany\\
$^{4}$ \quad Department of Physics, Geology, and Engineering Technology, Northern Kentucky University, 1 Nunn Drive, Highland Heights, KY      41099, USA\\
$^{5}$ \quad Guangxi Key Laboratory for Relativistic Astrophysics, School of Physical Science and Technology, Guangxi University, Nanning      530004, China\\
$^{6}$ \quad Astronomical Observatory Belgrade, Volgina 7, 11060 Belgrade, Serbia\\
$^{7}$ \quad Sorbonne  Universit\'e, CNRS/IN2P3, Laboratoire de Physique Nucl\'eaire et de Hautes Energies, LPNHE, 4 place Jussieu, 75005        Paris, France \\
$^{8}$ \quad Universit\'e Paris-Cit\'e, AstroParticules et Cosmologie, APC, 10 rue Alice Domon et L\'eonie Duquet, 75013 Paris, France\\
$^{9}$ \quad CEA-Saclay, IRFU/Departement d’Astrophysique, 91191 Gif-sur-Yvette, France\\
$^{10}$ \quad Institut d$^{\prime}$Astrophysique de Paris, Sorbonne Universit\'e, CNRS, UMR 7095, 98 bis bd Arago, 75014, Paris, France\\
$^{11}$ \quad Department of Astronomy and Physics, Saint Mary’s University, 923 Robie Street, Halifax, NS, B3H 3C3, Canada \\
$^{12}$ \quad Institut für Astrophysik und Geophysik, Universität G\"ottingen, Friedrich-Hund Platz 1, 37077 G\"ottingen, Germany \\
$^{13}$ \quad Ruhr University Bochum, Faculty of Physics and Astronomy, Astronomical Institute (AIRUB), 44780 Bochum, Germany\\
$^{14}$ \quad ESA, European Space Astronomy Centre (ESAC), 
Villanueva de la Ca\~{n}ada, E-28692 Madrid, Spain\\
}

\corres{Correspondence: dwxu@nao.cas.cn (D.X.); skomossa@mpifr.de, (S.K.)
}

\abstract{We discuss applications of the study of the new and barely explored class of changing-look (CL) narrow-line Seyfert 1 (NLS1) galaxies and comment on their detection with the space mission SVOM (Space Variable Objects Monitor). We highlight the case of NGC 1566, which is outstanding in many respects, for instance as one of the nearest known CL AGN undergoing exceptional outbursts. Its NLS1 nature is discussed, and we take it as a nearby prototype for systems that could be discovered and studied in the near future, including with SVOM.
Finally, we briefly examine the broader implications and applications of CL events in NLS1 galaxies and show that such systems, once discovered in larger numbers,
will greatly advance our understanding of the physics of the environment of rapidly growing supermassive black holes.
This White Paper is part of a sequence of publications which explore aspects of our understanding of (CL) NLS1 galaxy physics with future missions.
}


\begin{document}


\section{Introduction}

Highly variable and transient active galactic nuclei (AGN) provide us with important insights into the physics of the accretion of matter onto their central supermassive black holes (SMBHs). 
Of particular interest are changing-look (CL) AGN, which change their optical Seyfert-type (from type 1 to type 2, or vice versa) in response to high-amplitude continuum variability.
First recognized in the 1970s, such systems have been more recently identified in larger numbers based on dedicated searches for either strong variability in the optical or X-ray continuum emission, or alternatively in the broad emission lines \citep[see][for a review]{ KomossaGrupe2023}.
Several models have been suggested to explain CL AGN. Some explored extinction variability, for instance by dusty material crossing our line-of-sight or alternatively by dust destruction in response to an outburst \citep{Oknyansky2018}. The majority of theoretical models has focused on short-term changes in the accretion rate (on the time scale of months to decades), for instance due to disk instabilities or other mechanisms (e.g., the radiation-pressure disk instability, the Hydrogen ionization instability, magnetic pressure supported disks, magnetic flux inversion, instabilfities in warped disks at large radii, or shock formation in highly precessing disks)  \citep[e.g.][]{Nicastro2000, Janiuk2002, Grupe2015, Ross2018, NodaDone2018, DexterBegelman2019, Sniegowska2020, PanXin2021, RajNixon2021, Feng2021, Laha2022, Kaaz2023, Cao2023} or have explored the possibility of tidal disk interactions in binary SMBH systems \citep{WangBon2020}. 

The subgroup of AGN that has shown particularly high amplitudes of variability in the X-ray band are 
narrow-line Seyfert 1 (NLS1) galaxies \citep[review by][]{Gallo2018}. Therefore, the question is raised, how common the CL phenomenon is among NLS1 galaxies.
This subgroup of AGN is defined by their remarkable optical spectroscopic emission-line characteristics: the narrow widths of their broad Balmer lines from the broad-line region (BLR) with FWHM(H$\beta$) $<$ 2000 \kms\, and an emission-line intensity ratio of \oiii$\lambda5007$/H$\beta_{\rm total} < 3$  \citep{OsterbrockPogge1985, Goodrich1989}. 
The presence of \feii\, complexes is often added as additional defining criterion \citep{Veron2001}.
A number of NLS1 galaxies also show strong high-ionization forbidden lines in their optical spectra. 

NLS1 galaxies are of great interest because they are at one extreme end of AGN correlation space{\footnote{In a different AGN classification scheme, they are referred to as `population A' AGN, and the FWHM cut is set around 4000 km s$^{-1}$ \citep{Marziani2018}.}}, and trace AGN physics under extreme conditions (e.g., at low SMBH masses and high (near-Eddington) accretion rates).
Sky surveys have successfully provided us with large samples of  thousands of NLS1 galaxies. For instance, a large fraction of the soft X-ray AGN identified with the former space mission ROSAT turned out to be NLS1 galaxies \citep{Grupe1999, Xu2003}. More recently, large numbers of NLS1 galaxies were selected from the Sloan Digital Sky Survey (SDSS) \citep{Zhou2006, Rakshit2020, Paliya2024}.

NLS1 galaxies, as a class, are characterized by small SMBH masses and accretion rates near the Eddington limit (\citep{Boroson2002, Grupe2004, Xu2012}; review by \citet{Komossa2018}). They therefore represent systems of rapidly growing SMBHs in the local universe. Given their near-Eddington accretion rates, they should be particularly prone to CL events under those mechanisms which operate close to Eddington.
Here, we discuss implications and applications of the CL phenomenon in NLS1 galaxies, and discuss the detectability of such systems with the upcoming space mission SVOM. We highlight the example of the nearby CL NLS1 galaxy NGC 1566.
We use a cosmology with 
$H_{\rm 0}$=70 km\,s$^{-1}$\,Mpc$^{-1}$, $\Omega_{\rm M}$=0.3 and 
$\Omega_{\Lambda}$=0.7
and the cosmology calculator of \citet{Wright2006}. (In the case of NGC 1566, this corresponds to a Hubble-flow distance of 21.5 Mpc.) 

\section{CL events in NLS1 galaxies}
\label{sec:known-systems}

Only few NLS1-type galaxies have so far been identified as CL AGN (Tab. 1){\footnote{We note that here we adopt throughout the optical classification of CL events which has been common in recent years. An alternative scheme looks at X-ray spectral variability, and refers to CLs as events where the X-ray spectrum changes from highly absorbed as in type 2 AGN to barely or unabsorbed as in type 1 AGN \citep{matt2003}.  
We do not discuss possible such X-ray absorption events in NLS1 galaxies. 
}}. 
This Section provides a brief overview of their properties, and critically discusses the need to distinguish from other phenomena like type II SNe in dense media and (very rare) stellar tidal disruption events.

The AGN IC 3599 (Zwicky\,159.034) harbored an extreme CL event, dramatically changing its optical emission-line spectrum following a high-amplitude soft X-ray (and optical) outburst with a peak luminosity of $L_{\rm x} = 10^{44}$ erg\,s$^{-1}$ 
\citep{Brandt1995, Grupe1995}.  Optical spectra taken during the first outburst revealed bright optical emission lines, including broad Balmer lines with FWHM of 1200 \kms\, and high-ionization iron lines of similar width \citep{Brandt1995}. These then faded away strongly in subsequent years. 
In low-state, IC 3599 is classified as Sy 1.9 galaxy with faint broad-line emission only detectable in H$\alpha$ and with a long-lived narrow-line region (NLR) including faint high-ionization iron lines \citep{KomossaBade1999}.  
While IC 3599 is sometimes referred  to as NLS1 galaxy based on its strong variability and its FWHM(H$\beta$) $\sim$1200 \kms, it is not a typical NLS1 because no \feii\, emission was detected at high-state,  and because all permitted and forbidden lines had the same width at high-state more typical of a coronal-line region (CLR). However, in any case, IC 3599 can certainly be regarded as an extreme case of a CL AGN. 

The high-amplitude outbursts of IC 3599 were suggested by \citet{Grupe2015} to be due to possible thermal instabilities in the accretion disk, propagating through the inner region at the sound speed and causing an increase of the
local accretion rate.  Episodes of repeat flaring are then produced on the timescale of decades when the inner disk empties and refills. The two bright outbursts of IC 3599 were separated by 20 yrs.  
Other scenarios to produce repeated photoionizing outbursts,  like an OJ 287-type binary SMBH scenario (where a secondary SMBH might interact in one way or another with the accretion disk around a primary SMBH),  remain possibilities \citep{Komossa2014}. 

\begin{table}[H]
\caption{CL events among candidate NLS1 galaxies. Column (1) provides the galaxy name, column (2) its redshift $z$, column (3) the total amplitude of variability in X-rays (or in the optical band if no X-ray observations were available), column (4) the FWHM of the broad H$\beta$ emission line, 
and column (5) the variability time scale $\Delta$t$_{\rm var}$ between the optical spectroscopic change (from one type to another). Since often only two spectra exist, not a dense monitoring, it is well possible that the AGN changed its type multiple times in between, but simply no spectra were taken to record this. $\Delta$t$_{\rm var}$ therefore usually is an upper limit.  NGC 1566 changed within months from one type to the other, and here $\Delta$t$_{\rm var}$ of the fastest recorded change is listed.
Column (6) lists further comments. For references on the source parameters see Sect. 2.}\label{tab:data}
  \begin{adjustwidth}{-\extralength}{0cm} 
\centering
\begin{tabularx}{\fulllength}{lcccll} 
\toprule
\\
name & $z$ & amplitude of var.& FWHM(H$\beta$)  &  $\Delta$t$_{\rm var}$ & comments \\
      &    & in X-rays & in \kms\, & in yrs &  \\
(1) & (2) & (3) & (4) & (5) & (6)\\
      \\
\midrule
\\
IC 3599 & 0.021 & $>$100 & $\sim$1200 & 0.75 &  two outbursts separated by 20 yrs \\
 & &  &  & & no bona-fide NLS1, no \feii \\ 
SDSS J123359.12+084211.5 & 0.256 & $\sim 2$  (opt) & 2430 & 11 & change in both Balmer and \feii\, emission \\ 
ZTF18aajupnt & 0.037 &  10 & 940 & 16& spectral change from LINER to NLS1 \\  
J1406507--244250  & 0.046 & $\sim$ 1 (opt) & 3000  & 18 & little optical continuum variability\\ 
NGC 1566 & 0.005 & 30 & 1950 & 0.33 &  repeat Seyfert-type changes \\
\\
\bottomrule
\end{tabularx} 
\end{adjustwidth} 
\end{table}

SDSS J123359.12+084211.5 was identified as CL NLS1 \citep{MacLeod2019} with changes in Balmer lines and particularly in \feii\, emission. In high-state, its FWHM (H$\beta$) was 2430 \kms\, \citep{Liu2019}. The continuum itself varied by a factor $\sim$ 2. 

ZTF18aajupnt was classified as NLS1 in high-state, with a LINER spectrum in its low-state \citep{Frederick2019}.  
Its high-state spectrum shows strong coronal lines. 
A luminous soft X-ray and mid-infrared flare was detected.

J1406507--244250  was identified from 6dFGS as a candidate CL NLS1 \citep{Hon2022}. Its optical continuum is about constant, but its H$\beta$ line was found to change its FWHM from 500 \kms\, to 3000 \kms, leading the authors to speculate about a possible NLS1 classification, taking the FWHM classification criterion loosely.   

   \begin{figure}[t]
   \centering
  \includegraphics[width=9cm]{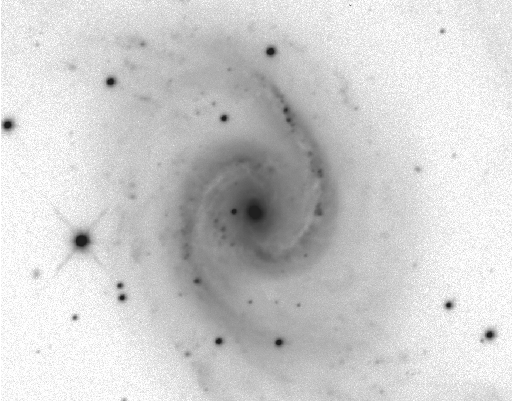}
   \caption{Optical image of NGC 1566 taken on 2023 October 19 with the LCOGT (Las Campanas Observatory Global Telescope) 1\,m telescope
   at Siding Spring Observatory. North is up, East is left.}
              \label{fig:opt_ima}%
    \end{figure}

There are also CL impostors. 
Some other events with transient broad lines have been identified in AGN, but their nature remained uncertain.  
This is because two other transient phenomena are accompanied by temporary emission of optical continuum and broad lines, therefore potentially mimicking a CL AGN. The first transients are supernovae (SN) of type IIn \citep{Filippenko1997} which show temporary broad and narrow emission lines (which are believed to arise in interactions of the SN ejecta with dense circum-stellar matter),
the other transients are optical stellar tidal disruption events \citep[TDEs;][]{Rees1988}. These are often accompanied by broad emission lines of the Balmer series and/or \textrm{He}\,{\footnotesize II}
\citep{Komossa2008, Onori2022} thought to arise from the temporary accretion disk \citep{Liu2017}. Some TDEs also show transient high-ionization coronal lines \citep{Komossa2008, Wang2012, Short2023}.  If TDEs occur in quiescent, in-active host galaxies, they can be identified from the host galaxy spectra showing inactivity before the transient event and/or long after the transient event. 
However the situation is more complicated when the galaxy is an AGN, such that it is not immediately clear, if the change comes from an accretion disk (instability) or, less likely, rather from a (very rare) TDE, or an SN IIn. The situation is also challenging, when there is no pre-outburst spectrum available, or only a few post-outburst spectra, so that long-lived AGN, rare TDE, or SN IIn scenarios cannot immediately be distinguished on optical spectra alone [see the detailed discussion of \citet{Komossa2008} and  \citet{Komossa2009}; and see \citet{Drake2011}, \citet{Blanchard2017} and \citet{Zhang2022} for examples in NLS1-like galaxies]. 

TDEs are distinguished by their supersoft X-ray emission \citep{KomossaBade1999, Li2020, Hampel2022}, except for jetted TDEs which are associated with radio emission and have much harder X-ray spectra \citep{Burrows2011, Bloom2011} and could therefore be directly triggered with the ECLAIRs instrument aboard SVOM \citep{Wei2016}. 
SN IIn, on the other hand, are distinguished by the absence of (long-lived) bright X-ray emission, or the absence of X-rays altogether.  Further, TDEs are very rare events with a rate of 10$^{-4}$--10$^{-5}$ events per galaxy per year \citep{WangMerritt2004}, and we therefore would not expect any significant number of these in the most nearby galaxies, if any at all.  
While TDEs themselves are another interesting source population for SVOM, these will not be discussed further here. 
Many will be detected with the 
X-ray satellite Einstein Probe 
\citep{Yuan2016}, a Chinese mission with European participation which was launched in January 2024.

Why have so few CL events been so far observed in NLS1 galaxies?
Balmer lines in NLS1 galaxies are bright because of high accretion rates near the Eddington limit, in NLS1 galaxies as a class. Therefore, a {\em strong} change in BLR luminosity is required before a system is identified as CL event (instead, if the broad lines are already faint in the first place, their disappearance and therefore a change to the type-2 state, would already be much more easily detectable). Ideally, CL events should therefore require a threshold amplitude of variability in the Balmer lines and continuum flux of a factor of, e.g., $>$ 10, in order
to identify CL systems homogeneously among AGN. 
Otherwise, AGN with faint (broad) emission lines would be preferentially classified as CL systems. 

   \begin{figure}[t]
   \centering
  \includegraphics[clip, trim=0.9cm 5.3cm 1.0cm 6.3cm, width=13cm]{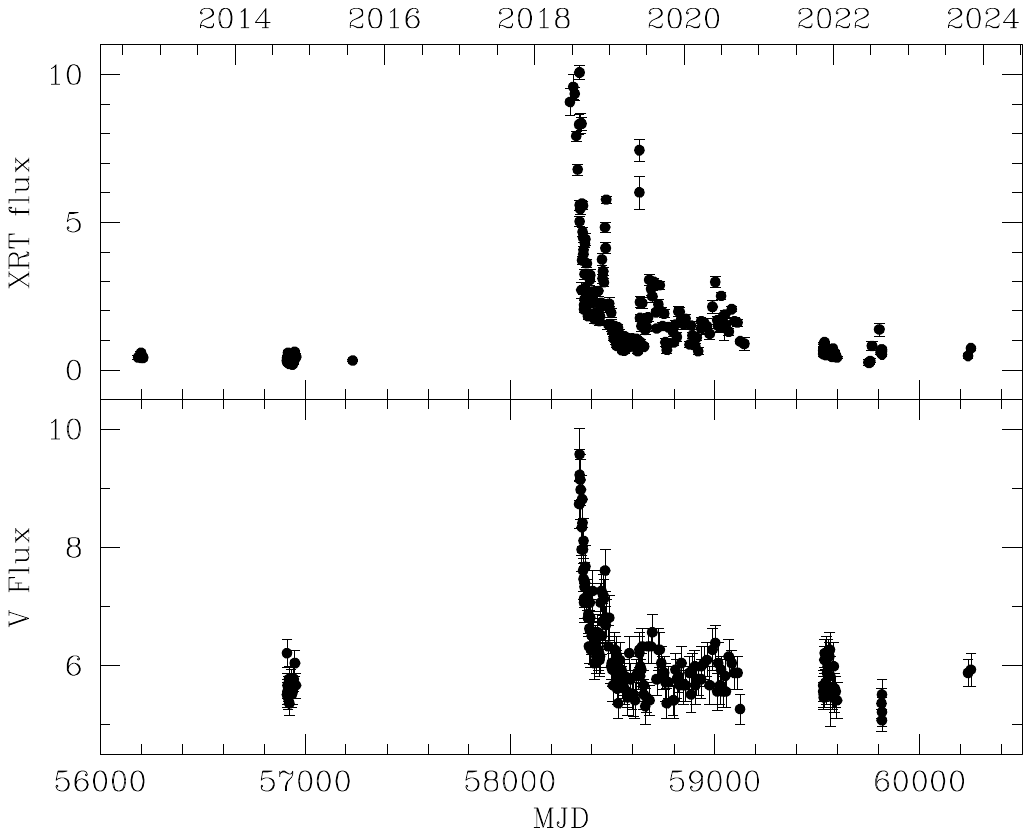}
   \caption{Swift lightcurve of NGC 1566. 
   Upper panel: X-ray flux in units of 10$^{-11}$ erg\,cm$^{-2}$\,s$^{-1}$ (0.3--10 keV), 
lower panel: optical V flux 
in units of 10$^{-11}$ erg\,cm$^{-2}$\,s$^{-1}$.
The latest bright outburst occurred in 2018 and was detected in all Swift bands. The last two data points are from 2023 October 19 and November 2, and show NGC 1566 in quiescence. }
\label{fig:lc_swift}%
    \end{figure}
    
\begin{figure}[t]
   \centering
  \includegraphics[clip, trim=0.9cm 5.3cm 1.0cm 13.3cm, width=12.0cm]{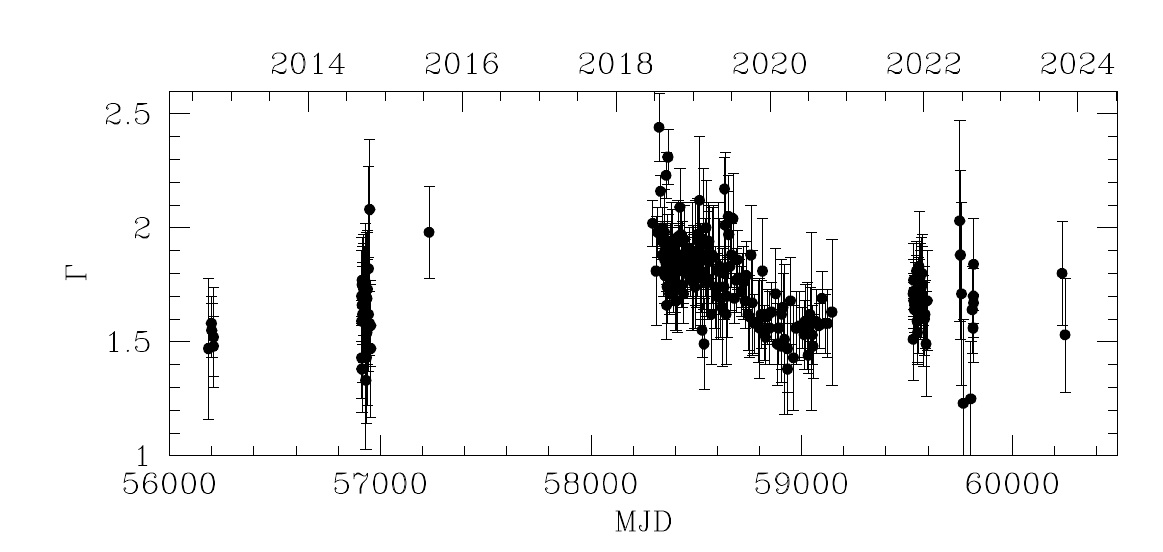}
   \caption{
   The variation of the X-ray photon index $\Gamma_{\rm X}$ of NGC 1566 from single-component powerlaw fits with absorption fixed at the Galactic value.}   \label{fig:gamma_swift}
    \end{figure}

Finally, the particularly interesting case of CL events in NGC 1566 will be further discussed in the next Section, including new data taken in 2023 October--November. 

\section{The NLS1 nature of NGC 1566 and the feasibility of detecting CL NLS1 galaxies with SVOM}

NGC 1566 (Fig. \ref{fig:opt_ima}) is the nearest known CL AGN {\em repeatedly} changing its Seyfert type.{\footnote{Even though this contribution is focused on NLS1 galaxies, we note in passing that some broad-line Seyfert 1 galaxies which changed their Seyfert-subtype are comparably nearby as NGC 1566. One such example is NGC 4151. Traditionally known as intermediate-type Seyfert galaxy \citep[e.g.][]{Seyfert1943}, its broad Balmer lines albeit still detected \citep{Penston1984, Kielkopf1985} became significantly fainter in 1984 \citep{Penston1984, Lyuty1984, Kielkopf1985} and were already back to their bright state in January 1985 \citep{Peterson1985}.
}}
Here, we argue for its NLS1 classification.
Initially known as a bright type 1 AGN \citep{Shobbrook1966}, it became much fainter in subsequent years.
Multiple spectroscopic observations between 1980 and 1982 found it to be a type 2 Seyfert in 1980, then gradually changing back to a type 1 Seyfert, and another CL event happened in 1985 \citep{Alloin1985, Alloin1986}. Most of the changes occurred within only 4 months. The Balmer decrement of the broad lines did not change significantly during the CL event \citep{Alloin1985}.

In 2018, an INTEGRAL detection of NGC 1566 at hard X-rays \citep{Ducci2018} triggered multi-wavelength (MWL) follow-up observations \citep{Ferrigno2018, Grupe2018, Dai2018, Oknyansky2018b}, 
revealing high-amplitude variability. These led to the identification of a new CL event in NGC 1566 \citep{Parker2019, Oknyansky2019, Ochmann2020}.

The width of the broad H$\beta$ line in NGC 1566 is relatively narrow. \citet{DaSilva2017} measured FWHM(H$\beta$) = 1950$\pm$20 \kms\, and FWHM(H$\alpha$) = 1970$\pm$10 \kms\, \citep[see also][]{Alloin1985, Ochmann2023}. Together with a small ratio of \oiii$\lambda5007$/H$\beta_{\rm total} < 3$, NGC 1566 fulfills the classification of a NLS1 galaxy. 
\feii\, emission is detected in the type 1 and 1.5 spectral state \citep{Ochmann2023}.
This makes NGC 1566 the nearest of the few CL cases in NLS1 galaxies detected so far.

The X-ray spectrum of NGC 1566 has a strong hard X-ray component, enabling its detection in high-energy surveys in outburst, as was the case with INTEGRAL in 2018.
\citet{Parker2019} reported the presence of a typical type 1 X-ray spectrum at high state, extending to high energies 
and subject to ionized absorption and mild reflection,  based on rapid X-ray follow-up observations carried out with XMM-Newton, NuSTAR and the Neil Gehrels Swift observatory (Swift hereafter). An outflow with velocity of 500 \kms, possibly launched by the outburst, was detected with the reflection grating spectrometer (RGS) of XMM-Newton. 
An earlier hard X-ray flare is evident in 2010 in the Swift BAT lightcurve \citep{Oh2018}.

The Swift longterm optical and X-ray lightcurve is displayed in Figs. \ref{fig:lc_swift} (V band and X-ray flux) and \ref{fig:gamma_swift} (X-ray spectral shape, parameterized by a power law). The majority of observations was proposed by (some of) us, with the addition of archival data including serendipitous observations in 2022 due to an SN program. Our most recent measurements of NGC 1566 of 2023 October 19 and November 2 
show it to be in its quiescent state. 

The Swift \citep{Gehrels2004} data reduction was carried out following standard procedures. In brief, the UV-optical telescope \citep[UVOT;][]{Roming2005} observes in three optical and three UV filters. The UV and optical bands in NGC 1566 are closely correlated and we representatively show the V band fluxes here. The V filter central wavelength is $\lambda$=5468\AA.  For further analysis, the data sets of each V-band observation were first co-added. Source counts were then extracted in a region of circular size with an extraction radius of 3 arcseconds centered on NGC 1566.  A nearby area of 20 arcseconds radius was used to extract the  
background region. The  
background-corrected counts were converted into VEGA magnitudes and 
flux densities using the latest calibration (the flux densities in each filter were
 determined with the UVOT F-tool {\it uvotsource} with the parameter {\it apercorr} set to {\it curveofgrowth} in order to compensate for the deviation from the calibrated circular 5 arcsecond radius source extraction region; the smaller 3 arcsecond extraction region was chosen here to minimize the contribution of the host galaxy).
The V-band fluxes are reported as flux density multiplied by the central frequency of the V filter.
The last two UVOT data points of 2023 are affected by a slight, uncorrected drift motion of the satellite such that sources are no longer pointlike. Therefore, an elliptical extraction region was chosen instead. Finally, the data were corrected for Galactic reddening
based on the reddening curves of \citet{Cardelli1989}.
The  X-ray data were obtained with the Swift X-ray telescope \citep[XRT;][]{Burrows2005}. To carry out the timing and spectral analysis, we selected source photons within a circular area with a radius of 20 detector pixels, where one pixel corresponds to a scale of 2.36 arcseconds. Background photons were extracted in a nearby circular region. X-ray spectra of the source and background in the energy band 0.3-10 keV were generated. The spectral analysis was then carried out with the package XSPEC \citep{Arnaud1996}.                                                

\begin{figure}[t]
   \centering
   \includegraphics[width=10cm]{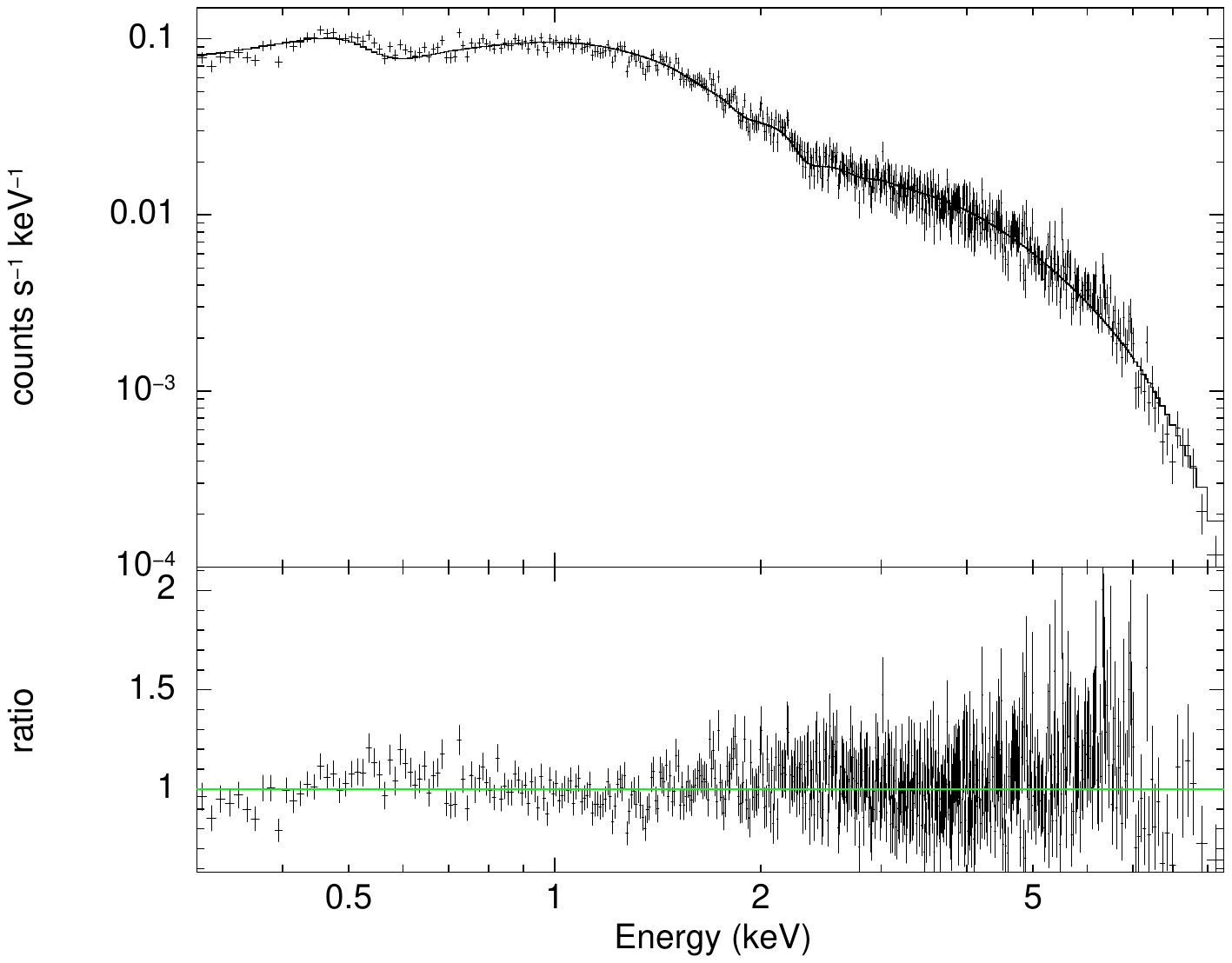}
   \caption{Composite Swift X-ray spectrum of NGC 1566 created from all intermediate to low-state observations taken in the PC (photon counting) mode, with a total exposure time of 264 ks, overbinned and fit with a single powerlaw spectrum. 
   }
              \label{fig:spec_swift}%
    \end{figure}

In order to measure its average X-ray spectral slope, we have merged all existing intermediate to low-state Swift observations of NGC 1566 into one single deep spectrum (amounting to a total exposure time of 0.26 Megaseconds) using the software tool of \citet{Evans2007}. The spectrum was then fit with a single powerlaw subject to Galactic absorption ($N_{\rm H} = 0.9 \times 10^{20}$ cm$^{-2}$). This results in an average powerlaw photon index of $\Gamma$ = 1.6 (Fig. \ref{fig:spec_swift}) and demonstrates the rather hard X-ray spectrum of NGC\,1566. 

CL systems like NGC 1566 will be important `secondary science' targets for the SVOM mission.
SVOM (Space Variable Objects Monitor) is a Sino-French space mission with international participation that is scheduled to launch in June 2024 \citep{Wei2016} with four instruments onboard: two gamma-ray telescopes 
\citep[ECLAIRs (4--150 keV) 
and GRM (15--5000 keV);][]
{Godet2014, Godet2022,He2020}, 
one narrow-field X-ray telescope \citep[MXT; 0.2--10 keV;][]{Gotz2014, Gotz2023}, 
and an optical telescope \citep[VT; 400--650 nm and 650--1000 nm;][]{Wang2020}. 
The field of view of each detector is 2 sr (ECLAIRs and GRM), $64\times64$ arcmin$^2$ (MXT) and $26\times26$ arcmin$^2$ (VT), respectively. The satellite will be placed into a LEO (Low Earth Orbit) with an inclination of 30 degrees, an altitude of 625 kilometers, and an orbital period of approximately 96 minutes.
Its primary goal is the discovery and study of GRBs at all redshifts \citep{Wei2016,Dagoneau2020,Yu2020}. 
However, it will also enable a great wealth of other transient astrophysics
which is driven by proposals. The Target of Opportunity proposals are open to all scientists and are evaluated by the Principal Investigators.
Here, we comment on the detection of CL NLS1 galaxies with SVOM, highlighting the galaxy NGC 1566 as a nearby prototype. 

First, the initial trigger of an NGC 1566 like outburst would come from ECLAIRs. 
During the SWIFT BAT survey, NGC 1566 was detected at an average peak flux of 4 mCrab (or 10$^{-10}$\,erg\,cm$^{-2}$\,s$^{-1}$)
in a one month interval during its 2010 outburst \citep{Oh2018} in the BAT energy band of 14--195 keV with a powerlaw photon index $\Gamma$ = 2.2. 
The sensitivity limit of ECLAIRs is 7.9 $\times$ 10$^{-10}$\,erg\,cm$^{-2}$\,s$^{-1}$
for a 5 sigma detection in 1 ks \citep{Paul2011,Godet2014,Cordier2015}.
During a new outburst, and at a flux level of 4 mCrab, NGC 1566 would be detected with ECLAIRs within 17 ks at 5 sigma.  
NGC 1566 would therefore be well above the detection threshold of ECLAIRs.

With the Swift XRT, NGC\,1566 was not monitored during the 2010 outburst, but well during its 2018 outburst (Fig. \ref{fig:lc_swift}). Its (0.3--10 keV) peak flux was f = 10$^{-10}$\,erg\,cm$^{-2}$\,s$^{-1}$
\citep{Parker2019}. A similar outburst would therefore be detected at 5 sigma with the SVOM MXT \citep{Gotz2014} with an integration time of 16 s.   
The detection limit of the VT is 22.5 mag in a 300 s observation, and systems like NGC 1566, or even much fainter, would be easily observable and followed-up with the VT, even in their low-emission state. Because of its two filters, VT also provides colour information. 

In general, AGN outbursts with an (isotropic) X-ray peak luminosity of 10$^{43}$ erg\,s$^{-1}$ will be detectable out to 200 Mpc with MXT in 10 ks. 

\section{Implications and applications of studying CL events in NLS1 galaxies}

In addition to investigating the physics of the BLR in NLS1 galaxies and measuring the cross-band time delays, identification and follow-ups of CL events in NLS1 galaxies 
have other important astrophysical applications.

First, the disk-instability models make CL NLS1 galaxies valuable targets in shedding light on the type and origin of the disk instability and the accretion disk physics in general under extreme physical conditions. 
By adopting values of the viscosity parameter $\alpha$ and the radiative efficiency parameter $\eta$ of $\alpha=0.1$ and $\eta=0.1$, the timescales are estimated to be $\sim10$ and $\sim6500$ yrs for a $10^7\ M_\odot$ SMBH accreting at the Eddington limit in the thermal and viscous radial inflow scenarios \citep[e.g.,][]{Siemiginowska1996, LaMassa2015, Lawrence2018}, respectively.
Additional mechanisms, such as elevation of the accretion disk by a magnetic field 
\citep[e.g.][]{DexterBegelman2019, PanXin2021}, or any of the other mechanisms discussed in the literature,
may therefore be required to understand  shorter timescales.

Second, the search for a population of type 2 NLS galaxies, if they exist \citep{Dewangan2005}, has been difficult to address, since the main NLS1 identification relies on the BLR properties (line width, \feii\, strength, \oiii$\lambda5007$/H$\beta_{\rm total}$ ratio), unobservable in type-2 states. 
Few candidate type 2 NLS galaxies have therefore been identified so far, relying mostly on polarization observations \citep[e.g.,][]{Pan2019, Pan2021}.
Once identified (no matter whether the type 2 state is long-lived and due to the unified model \citep{Antonucci1993}, or due to high-amplitude variability), then, in X-rays, type 2 states enable follow-up imaging spectroscopy:
e.g., detection of X-ray lines from the circum-nuclear medium \citep{Tomas2022, Buhariwalla2023} and determination of their excitation/ionization mechanism, or detection of optical outflows spatially resolved, of great interest for our understanding of mechanisms of feeding and feedback, especially in these high-accretion systems rapidly growing their SMBHs.  

Third, optical stellar velocity dispersion measurements become more accessible in type 2 states, otherwise known to be particularly challenging in NLS1 galaxies 
with their bright continuum emission and strong \feii\, complexes in their type 1 states. 

Fourth, type 2 states will allow us to study the locus of NLS1 galaxies in diagnostic diagrams \citep{Veilleux1987}. These are  based on narrow emission-line ratios as diagnostic tools of the physics of the NLR and of the EUV emission of the accretion disk. 
This has so far rarely been done because of the challenge of deblending broad and narrow lines when the broad-line FWHM is so small ($<$2000 \kms). Samples of at least 10--20 type 2 NLS galaxies will be required for a systematic study of their location in diagnostic diagrams. Along with comparative studies with other types of AGN, especially broad-line Seyfert 1 (BLS1) galaxies, this will provide valuable insights into understanding the characteristics of the NLR properties in NLS1 galaxies.

Fifth, as hypothesized "young" AGN in the early stages of evolution, CL NLS1 galaxies provide an ideal platform to examine the co-evolution between AGN and their host galaxies \citep{Heckman2014}. This can be done by directly investigating the properties of their host galaxies through their type 2 state. Such studies have been hindered due to the strong nonstellar radiation from the luminous active nucleus, which typically overwhelms the starlight from the host galaxy.

Sixth, the stellar populations of the host galaxy can be easily measured in type 2 states. Studies of the stellar population of the host galaxy can give direct evidence if type 2 NLS galaxies possess young population. 

Finally, while the focus of this publication is on using the CL phenomenon for new applications to understand NLS1 physics, we also note that systematic comparisons between CL BLS1 and CL NLS1 galaxies will highlight possible systematic differences in their accretion disk properties (for instance, different types of disk instabilities operating at different $L/L_{\rm Edd}$).

\section{Summary and conclusions}
The space mission SVOM, scheduled to be launched in 2024, will detect nearby CL AGN. Here, we have focused on NLS1 galaxies in particular, have discussed multiple new applications of observing NLS galaxies in their type 2 state, and have highlighted NGC\,1566 as a nearby CL prototype, that showed bright MWL outbursts in the past 
and is found in a quiescent state in our most recent Swift observations.
New hard X-ray transients will first be triggered with ECLAIRs.  
The MXT with its soft X-ray sensitivity, and the highly sensitive VT, are then particularly well suited for follow-ups of all bright CL AGN, NLS1 and BLS1 galaxies. 
In addition to serving as the discovery mission, SVOM can also be used to follow-up CL events first detected in other surveys and/or wavebands. 

The identification of further CL AGN, and in particular more CL NLS1 galaxies, is also expected from large optical spectroscopic surveys with repeat spectroscopic coverage, like SDSS \citep{MacLeod2016, Runco2016},
 the Large Sky Area Multi-Object Fiber Spectroscopic Telescope \citep[LAMOST;][]{Yang2018} 
or the Dark Energy Spectroscopic Instrument 
 \citep[DESI;][]{Guo2023}, 
and will continue to be extremely valuable.   
This approach, however, will generally miss the detection of new events as they happen, when they are brightest, so that MWL observations will not be triggered quasi-simultaneously. 
In contrast, the identification of transient events among AGN with SVOM will enable {\em rapid} follow-up observations including spectroscopy. SVOM will issue alerts on bright transients within $\sim$5 minutes \citep{Wei2016}.  

In addition to their identification and X-ray and optical follow-ups with the instruments onboard SVOM to follow the lightcurve evolution, triggered deeper IR, optical and X-ray imaging spectroscopy will then allow important new applications, especially w.r.t. the study of type 2 NLS galaxies, their NLR properties, their circum-nuclear environment, and of mechanisms of feeding and feedback. This will provide new insights and constraints on the formation and evolution of SMBHs.



\vspace{6pt} 




\authorcontributions{
D. X. and S. K. developed the initial  idea and drafted the first version of the manuscript.
All authors discussed about changing-look NLS1 galaxies and their detection with SVOM, commented on and contributed to the manuscript. 
All authors have read and agreed to the published version of the manuscript.
}

\funding{This research was funded by the National Natural Science Foundation of China of grant numbers 12273054 and 12173009. 
EB would like to acknowledge the support of the Serbian Ministry of Education, Science and Technological Development, through the contract number 451-03-68/2023-14/200002. MWO gratefully acknowledges the support of the German Aerospace Center (DLR) within the framework of the `Verbundforschung Astronomie und Astrophysik' through grant 50OR2305 with funds from the German Federal Ministry for Economic Affairs and Climate Action (BMWK).
}


\dataavailability{
All data are available upon reasonable request. The Swift data are available in the Swift archive at \url{https://swift.gsfc.nasa.gov/archive/}. 
} 



\acknowledgments{
 We would like to thank the Swift team for carrying out the observations we proposed. Additional Swift observations were taken from the archive. 
 We would like to thank Alexis Coleiro for very useful discussions, and our referees for their very useful comments. 
 This work made use of data supplied by the UK Swift Science Data Centre at the University of Leicester.
 It is our pleasure to thank the organizers and participants of the conference on "Astroinformatics and Virtual Observatories" (Renhuai, 2023 October) for an excellent meeting and discussions. 
 }

\conflictsofinterest{The authors declare no conflict of interest.
} 





\begin{adjustwidth}{-\extralength}{0cm}

\reftitle{References}


\bibliography{bibtex}


%


\PublishersNote{}
\end{adjustwidth}
\end{document}